\documentstyle[12pt,epsfig]{article}
\newcommand{\be}{\begin{equation}}
\newcommand{\ee}{\end{equation}}
\newcommand{\ba}{\begin{eqnarray}}
\newcommand{\ea}{\end{eqnarray}}
\newcommand{\bb}{\begin{thebibliography}}
\newcommand{\eb}{\end{thebibliography}}
\newcommand{\ci}[1]{\cite{#1}}
\newcommand{\bi}[1]{\bibitem{#1}}
\newcommand{\lab}[1]{\label{#1}}

\begin{document}
\phantom{.}
\vspace{.7cm}
\begin{center}
{\large{Spin  Effects  In Diffractive
 High-Energy Reactions}}
\\
S.V.Goloskokov,
%\footnote{Email: goloskkv@thsun1.jinr.dubna.su}
\\
Bogoliubov Laboratory of Theoretical  Physics,\\
 Joint Institute for Nuclear Research,\\
Dubna 141980, Moscow region, Russia
\end{center}

%\vspace{.7cm}
\begin{abstract}
Spin effects in diffractive $pp$ and $lp$ reactions are analyzed.
It is shown that the structure of the pomeron coupling can be
studied in future polarized experiments.
\end{abstract}

  Diffractive processes at high energies have common
properties. They lead to the events with large rapidity gap
between produced particles. The observation of these events at
CERN and DESY \cite{ua8,h1} has revived  interest in the
study of the pomeron and
reactions where its properties can be investigated. Future
 experiments at RHIC and HERA \cite{future,gur,compass} and
proposed
FELIX experiment at LHC \ci{felix} will permit one to study
different aspects of polarized diffractive $pp$ and $lp$
reactions. In some experiments, the diffractive
contributions might be an important background effect. All these
processes should carefully be analyzed to obtain  information
on what kind of dynamics at large distances can be
studied in future experiments and what effects in physical
observables should be expected. In this report we shall discuss
the effects
of spin-dependent pomeron coupling
in elastic $pp$ scattering and  diffractive $Q \bar Q$
leptoproduction. Similar problems have been discussed in
\ci{goladr}.

 The "standard" approaches \cite{pom} to the pomeron
exchange do not lead
to the spin-flip part in the pomeron coupling. However, some
model approaches predict nonzero spin effects in the $s \to
\infty, |t|/s \to 0 $ limit (see \ci{pum,soff,gol} e.g.).
This means that
the pomeron might not conserve the $s$-channel helicity.
The "new" property of the pomeron can provide very definite
predictions at high energies. This conclusion is  especially
important
for the diffractive scattering of polarized particles where the
complicated
spin structure of the pomeron should be manifested.

I shall discuss here some predictions of
the meson-cloud model (MCM) \cite{gol}. They now become actual
for the polarized PP2PP experiment at RHIC \cite{gur} where the
spin
asymmetries near the diffraction minimum should be
studied and for the proposed FELIX experiment at LHC  \ci{felix} where
elastic $pp$ scattering might be analyzed up to $|t| \sim 20
GeV^2$.

The MCM \cite{gol} provide the following form of
the pomeron-proton coupling:
\be
V_{ppI\hspace{-1.1mm}P}^{\mu}(p,r)=m p_{\mu} A(r)+ \gamma_{\mu}
B(r),
\lab{prver}
\ee
where $m$ is the proton mass and $r$ is the momentum transfer
($t=r^2$).
The $ \gamma_{\mu} B(r)$ term is a standard pomeron coupling
that determines the spin-non-flip amplitude. The term $m p_{\mu}
A(r)$ leads to the
spin-flip in the pomeron vertex which is caused by the
meson-cloud effects
and does not vanish in the $s \to \infty$ limit.
Really,  we can estimate
the Born terms of the
spin-non-flip and spin-flip effects from the pomeron-proton
vertex (\ref{prver})
\ba
|T^B_{++}(s,t)| \propto s \;|B(r)|; \nonumber\\
|T^B_{+-}(s,t)| \propto m \; \sqrt{|t|}\; s\; |A(r)|. \lab{fpm}
\ea
So, both the amplitudes have the same energy dependence.

The model predicts the following ratio for spin-flip
and non-flip amplitudes:
\be
\frac{|T^B_{+-}(s,t)|}{|T^B_{++}(s,t)|} \simeq
\frac{\;\;\;m\;\sqrt{|t|}\;|A(r)|}{|B(r)|} \simeq
  0.05 \div
0.07 \;\;\; {\rm for} \;\;\;  |t| \sim 0.5GeV^2  \lab{fr}
  \ee
that is consistent with the estimations of Ref. \ci{akch}

The spin-non-flip and spin-flip amplitudes in MCM have the eikonal
form
\ba
T_{++}(s,t)\propto i s \int \rho d\rho J_0(r \rho)
\left( 1-\exp{[2 i\chi_0(s,\rho)]} \right); \nonumber \\
T_{+-}(s,t)\propto i s \int \rho^2 d\rho J_1(r \rho)
\chi_1(s,\rho) \exp{[2 i\chi_0(s,\rho)]},
\ea
where
\ba
\chi_0(s,\rho) \propto \chi_0^{center}(s,\rho)+\frac{\sqrt{s}}{2}
\int dz \alpha^2(s,R);\nonumber\\
\chi_1(s,\rho)\propto \frac{d}{d \rho} \int dz \alpha(s,R).
\lab{chi}\ea
Here $\chi_0^{center}$ is determined by the $B$ term in
(\ref{prver}),
$\alpha(s,R)$ being proportional to the $A$ term in the coordinate
space.
In MCM this amplitude is determined by the large-distance
structure of
the hadron. Just the $\alpha(s,R)$ contribution defines the
spin-flip amplitude at high energies
and gives the predominant contribution to the
$\chi_0(s,\rho)$ eikonal phase at superhigh energies $\sqrt{s} >
100 GeV$ because of the coefficient growing as $\sqrt{s}$.
Thus, the MCM predicts the strong correlation between the
polarization
phenomena at high energies and differential cross section
behaviour
at superhigh energies.

%%%%%%%%%%%%%%%%%%%%%%%%%%%%%%%%%%%%%%%%%%%%%%%%%%%%%%%%%
 \vspace*{-.5cm}
\epsfxsize=11cm
\centerline{\epsfbox{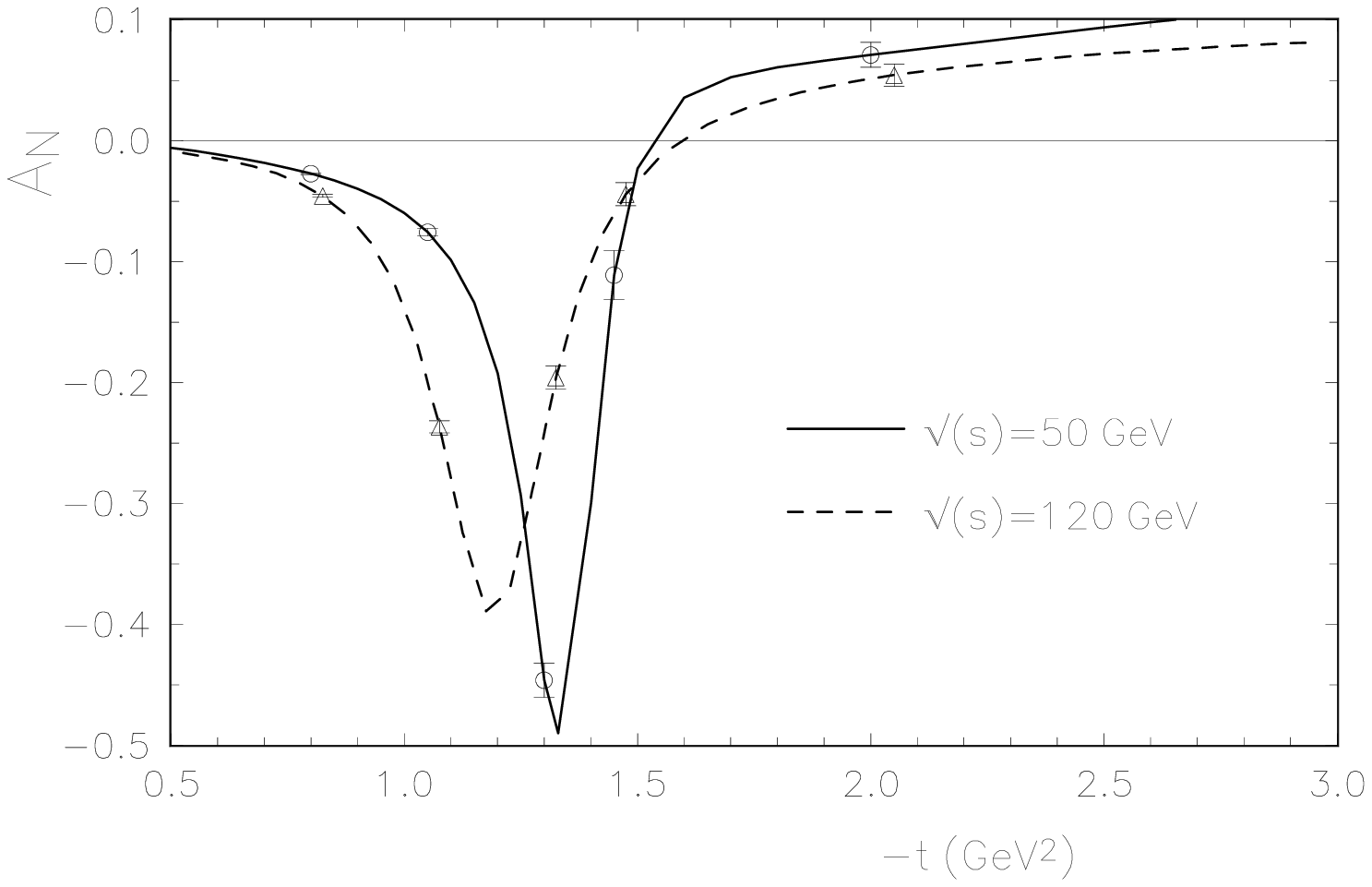}}
  \vspace*{.2cm}
\begin{center}
Fig.1
The predicted $A_N$ asymmetry of the $pp$-elastic scattering:
  full line -for  $\sqrt{s}= 50  GeV$;
   short-dashed line -for $\sqrt{s}= 120 \ GeV$.
The expected statistical errors are shown for PP2PP experiment.
\end{center}
%%%%%%%%%%%%%%%%%%%%%%%%%%%%%%%%%

The spin asymmetries are determined as a ratio of spin-dependent
and
spin-average cross section. Thus, spin asymmetries in PP
scattering
might be large near the diffraction minimum ($|t|
\sim
1GeV^2$) where the spin-average cross section is quite small.

The MCM model prediction for polarization at RHIC energies
with  the expected statistical errors for PP2PP
experiment \cite{gol97}
are shown in Fig.1. The expected errors are quite small and
the information about the spin-flip part of the proton-pomeron
coupling
can be obtained experimentally.

%%%%%%%%%%%%%%%%%%%%%%%%%%%%%%%%%%%%%%%%%%%%%%%%%%%%%%%%%
 \vspace*{-.5cm}
\epsfxsize=11cm
\centerline{\epsfbox{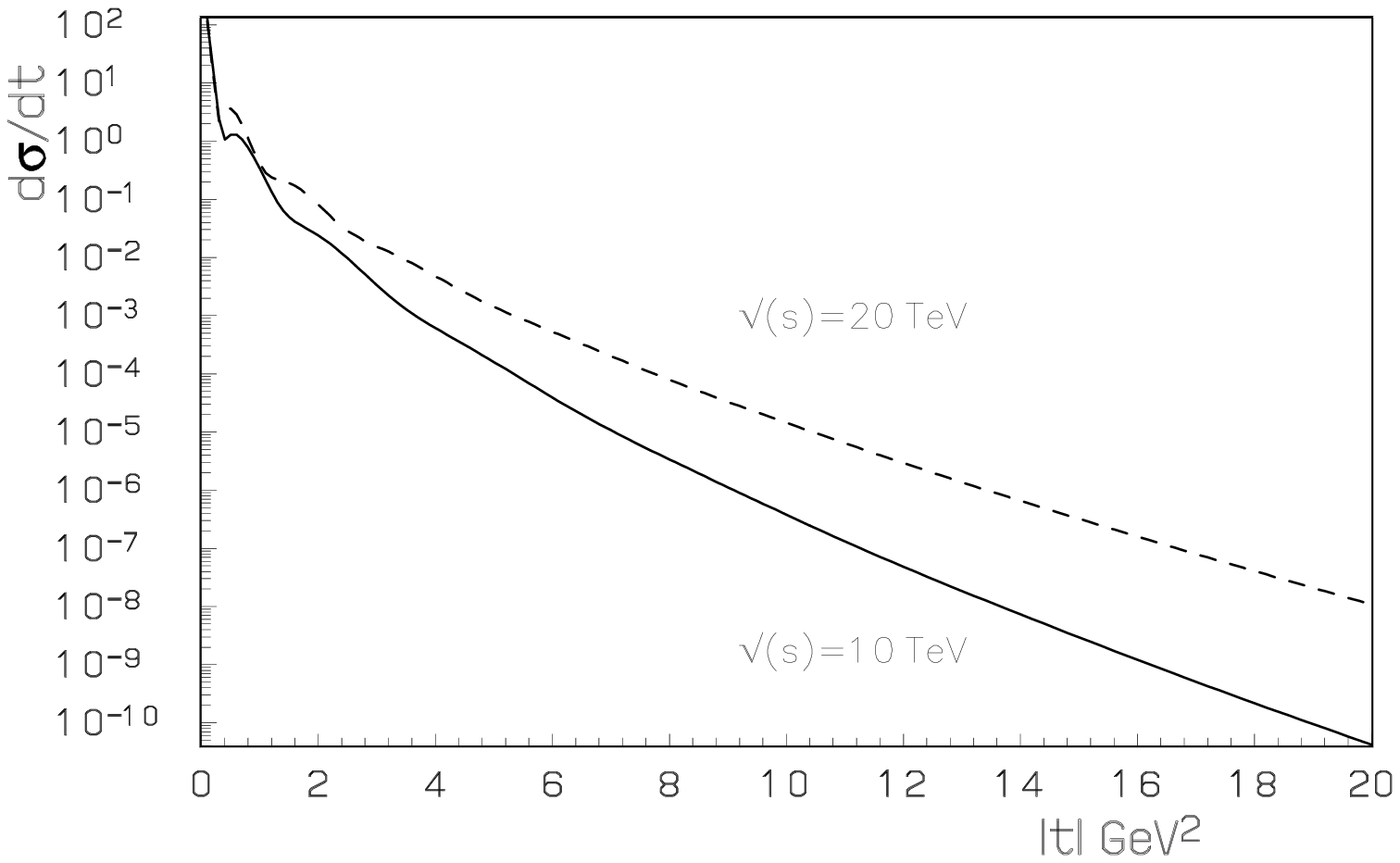}}
  \vspace*{.2cm}
\begin{center}
Fig.2
 Predictions for differantial cross section of pp - scattering:
       solid line -for  $\sqrt{s}=10 \ TeV$; dashed line -for
$\sqrt{s}=20 \  TeV$

\end{center}
%%%%%%%%%%%%%%%%%%%%%%%%%%%%%%%%%

  The model predictions for elastic proton-proton scattering at
LHC up to $|t| = 20 \ GeV^2$ are shown in Fig.2. The differential
cross section increasing at superhigh energies
is caused by the term  growing as $\sqrt{s}$    in the
eikonal phase $\chi_0$ (\ref{chi}).

So, the future PP2PP and FELIX experiments will give an excellent
possibility to test the spin structure of the pomeron-proton
coupling
and the possible rapid growth of the eikonal phase $\chi_0$
determined by the meson-cloud effects.

The future polarized diffractive experiments at
DESY \ci{future} and CERN \ci{compass} might give the
possibility
to study the spin structure of the pomeron in diffractive  lepton
proton reactions
\be
e+p \to e'+p'+X                          \lab{de}
\ee

It has been shown in \cite{golpl,golpr} that in addition to the
diagrams
where
gluons interact
with one quark in the hadron  \cite{pom}, the large-distance
gluon-loop effects
should complicate the structures of the pomeron coupling.
 The perturbative calculations \cite{golpl}
give the following form for this vertex:
\begin{equation}
V_{qqI\hspace{-1.1mm}P}^{\mu}(k,r)=\gamma^{\mu} u_0+2 M_Q k^{\mu}
u_1+
2 k^{\mu}
/ \hspace{-2.3mm} k u_2 + i u_3 \epsilon^{\mu\alpha\beta\rho}
k_\alpha r_\beta \gamma_\rho \gamma_5+i M_Q u_4
\sigma^{\mu\alpha} r_\alpha,    \label{ver}
\end{equation}
where $M_Q$ is the quark mass. We shall call the form (\ref{ver})
the spin-dependent pomeron coupling. If we consider only
the
$\gamma^{\mu} u_0$ contribution, it will be called  a standard
coupling.

Here we analyze the effects of the quark--pomeron
coupling  in the polarized diffractive $e+p \to e'+p'+Q \bar Q$
reaction based on \cite{golall,golhera,golcompass}
and estimate  the longitudinal
double--spin $A_{ll}$ asymmetry  for light and heave quark
production at energy $\sqrt{s}=20GeV$.

The difference of the polarized cross section  can be written
in the form
\ba
\Delta \sigma(t)=
\frac{d^5 \sigma(^{\rightarrow} _{\Leftarrow})}{dx dy dx_p dt
dk_\perp^2}-
\frac{d^5 \sigma(^{\rightarrow} _{\Rightarrow})}{dx dy dx_p dt
dk_\perp^2}= \nonumber\\
 \frac{3(2-y)\beta_0^4 F(t)^2 [9\sum_{i}e^2_i] \alpha^2}{128
x_p^{2\alpha_{P}(t)-1} Q^2 \pi^3} \frac{A(\beta,k_\perp^2,x_p,t)}
{\sqrt{1-4k_\perp^2\beta/Q^2}(k_\perp^2+M_Q^2)^2}. \lab{dsigma}
\ea
Here  $\sigma(^{\rightarrow} _{\Rightarrow})$ and
$\sigma(^{\rightarrow}
_{\Leftarrow})$ are the cross sections with parallel and
antiparallel
longitudinal polarization of the leptons and protons,
$\beta_0$ is the quark--pomeron coupling, $F(t)$
is the pomeron-proton form factor and $e_i$ are the quark charges.

The function $A$ is determined by the trace over the quark loop.
The contribution of the standard pomeron vertex to $A^s$
looks like
\be
 A^s(\beta,k_\perp^2,t)=
   16 (2 (1-\beta) k_\perp^2 - |t| \beta - 2 M_Q^2 (1+\beta))
|t|.
  \lab{ad}
 \ee
Similar forms can be written for the spin--average cross
sections.
In  both the cases  the strong dependence of the cross sections on
the mass of the produced quarks has been found.

We calculate the cross section
 integrated over momentum transfer because it is usually
difficult to
detect the recoil proton in diffractive experiments
\be
\sigma[\Delta \sigma]=\int_{t_m}^{0} dt
\sigma(t)[\Delta \sigma(t)],\;\;\;|t_m|=7(GeV)^2. \lab{intsi}
\ee
The exponential form of the proton form factor $F(t)=e^{bt}$
with
$b=1.9(GeV)^{-2}$ has been used.

%%%%%%%%%%%%%%%%%%%%%%%%%%%%%%%%%%%
  \vspace*{-.6cm}
%       \hspace*{.5cm}
\epsfxsize=11cm
\centerline{\epsfbox{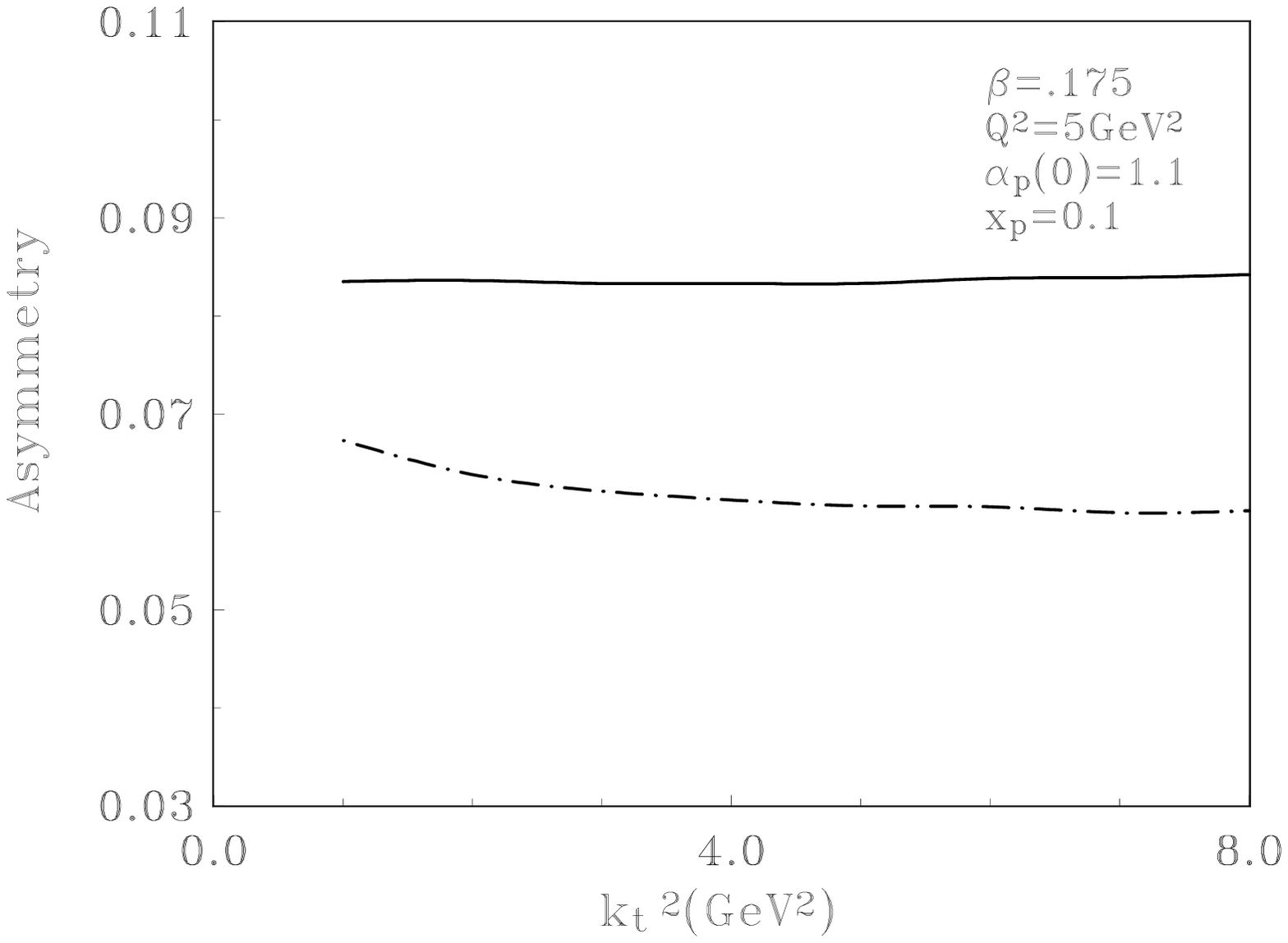}}
%{\epsfbox{gol_3.ps}}
  \vspace*{.3cm}
\begin{center}
Fig.3~$k^2_{\perp}$-- dependence of $A_{ll}$ asymmetry at
$\sqrt{s}=20(GeV)$.
Solid line -for the standard vertex;
dot-dashed line -for the spin-dependent quark-pomeron vertex.
\end{center}
%%%%%%%%%%%%%%%%%%%%%%%%%%%%====
The asymmetry of the diffractive light $Q \bar Q$ production is
shown in Fig. 3. It is sensitive to the spin structure of the
pomeron coupling.
The asymmetry for the standard quark--pomeron vertex is very
simple in form
\be
A_{ll}=\frac{y x_p (2-y)}{2-2y+y^2}.
\ee
There is no any $k_\perp$ and $\beta$ dependence here. For the
spin--dependent pomeron coupling the
 $A_{ll}$ asymmetry is smaller than for the standard
 pomeron vertex and depends on $k^2_{\perp}$. Thus, the
$A_{ll}$ asymmetry can be used to test the quark-pomeron coupling
structure.

The $A_{ll}$ asymmetry for diffractive open charm ($c \bar c$)
production has been
calculated for the standard pomeron coupling only.

%%%%%%%%%%%%%%%%%%%%%%%%%%%%%%%%%%%
  \vspace*{-.6cm}
%       \hspace*{.5cm}
\epsfxsize=11cm
\centerline{\epsfbox{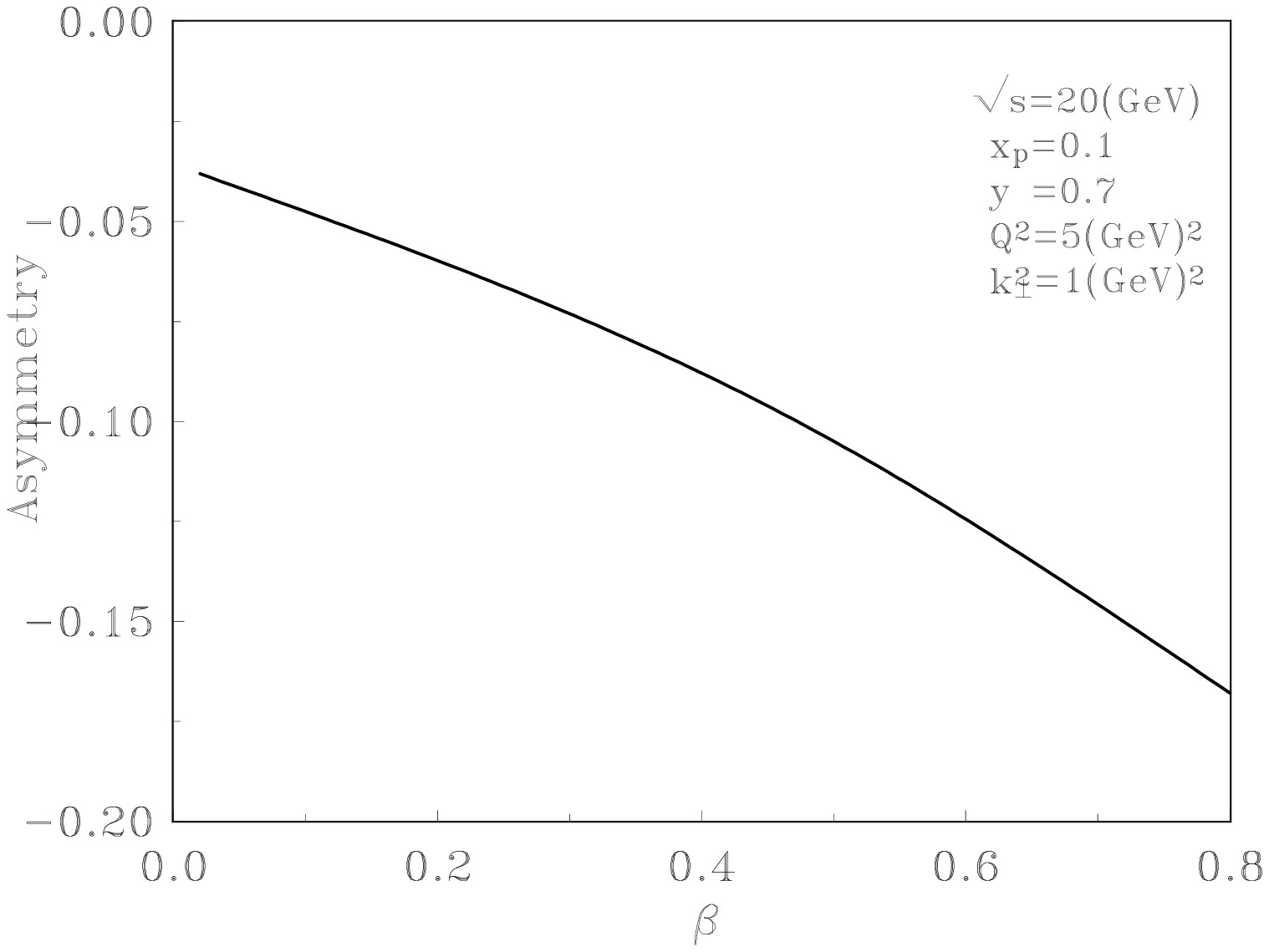}}
%{\epsfbox{asycompa.eps}}
  \vspace*{.3cm}
\begin{center}
Fig.4~$\beta$-- dependence of $A_{ll}$ asymmetry for diffractive
$c \bar c$ production at
$\sqrt{s}=20(GeV)$.
\end{center}
%%%%%%%%%%%%%%%%%%%%%%%%%%%%====
The predicted $A_{ll}$ asymmetry is not small. The open charm
prediction
is proposed to be used by COMPASS \ci{compass} to study $\Delta
G$.
It has been shown in
\ci{golcompass} that the produced diffractive $Q \bar Q$ jets
should be detected
by the COMPASS spectrometer. So, we can conclude that the
diffractive events might be an important background here.

We have found that the spin structure of the pomeron
coupling should modify the
spin average and  spin--dependent cross section in elastic and
diffractive
processes. It can be studied in single and double transverse spin
asymmetries
in elastic $pp$ scattering.
 The $A_{ll}$ asymmetry in diffractive $ Q \bar Q$
leptoproduction is
convenient to test the pomeron coupling
structure. The asymmetry is free from normalization
factors and is sensitive to the dynamics of pomeron interaction.
We have predicted large $A_{ll}$ asymmetry for the diffractive $Q
\bar Q$
production. This conclusion can be important in the analysis of
$\Delta G/G$
in the COMPASS  spectrometer.

Thus, we can conclude that the pomeron coupling structure
can be studied in different diffractive processes.
Note that the spin--structure of the pomeron vertex is
determined by
the large--distance gluon-loop correction or by the effects of
the hadron
wave function. So, the important test of the spin structure of
QCD at
large distances can be carried out by studying diffractive
reactions in future polarized experiments at HERA, RHIC and LHC
accelerators.

\end{document}